\documentclass[aps,prl twocolum,nofootinbib,superscriptaddress]{revtex4-2}
\usepackage{graphicx}
\usepackage{hyperref}
\usepackage{siunitx}
\usepackage{lineno}
\usepackage{graphicx}
\usepackage{amsmath}
\usepackage{color}
\usepackage{bm}
\usepackage{amssymb}
\usepackage{mathrsfs}
\usepackage{amsbsy}
\usepackage{caption}
\usepackage {ulem}
\usepackage[thicklines]{cancel}
\usepackage{diagbox}
\usepackage{subfigure}
\newcommand{\meqref}[1]{Eq. (\ref{#1})}
\newcommand{\mpref}[1]{Fig. \ref{#1}}

\newcommand{\be}{\begin{equation}}
\newcommand{\ee}{\end{equation}}
\newcommand{\bea}{\begin{eqnarray}}
\newcommand{\eea}{\end{eqnarray}}

\begin{document}
\title{Stationary equilibrium of test particles near charged black branes with the hyperscaling violating factor}

\author{Yu-Qi Lei}
\affiliation{Department of Physics, Shanghai University, 99 Shangda Road, Shanghai 200444, China}
\author{Xian-Hui Ge}
\email{Corresponding author: gexh@shu.edu.cn}
\affiliation{Department of Physics, Shanghai University, 99 Shangda Road, Shanghai 200444, China}

\begin{abstract}
We explore the upper bound of the Lyapunov exponent for test particles that maintain equilibrium in the radial direction near the charged black brane with the hyperscaling violating factor. The influences of black brane parameters (hyperscaling violation exponent $\theta$ and dynamical exponent $z$) are investigated. We show that the equilibrium in the radial direction of test particles can violate the chaos bound. The chaos bound is more easily violated for the near-extremal charged black branes. When the null energy condition ($T_{\mu\nu}\xi^\mu\xi^\nu \geq 0$) is broken, the bound is also more likely to be violated. These results indicate that the chaos bound of particle motion is related to the temperature of the black hole and the null energy condition (NEC). By considering the zero-temperature  and $T_{\mu\nu}\xi^\mu\xi^\nu=0$ cases, we obtain the critical parameters $\theta_c$ and $z_c$ for the violation of chaos bound. The chaos bound is always satisfied in the range $\theta > \theta_c$ or $z>z_c$.

\end{abstract}

\maketitle
\section{Introduction}
Chaos is an important nonlinear phenomenon that describes the sensitive response of the evolution of a system to the initial conditions. To illustrate the strength of chaotic phenomena, the Lyapunov exponent can be introduced. When a chaotic system is perturbated, the perturbation grows exponentially with time, and its corresponding exponent is the Lyapunov exponent. The larger the Lyapunov exponent, the more chaotic the test particle. As black hole theory is a kind of nonlinear theory, it is normal to see chaos in studying black holes. Chaos often exists in various objects, for example, the chaotic trajectories near black holes \cite{Suzuki:1996gm, Letelier:1997uv, deMoura:1999wf, Kao:2004qs, Chen:2016tmr, Wang:2016wcj, Ma:2014aha, Bera:2021lgw, Xie:2022yef} and the chaos in black hole thermodynamics \cite{Chabab:2018lzf, Mahish:2019tgv, Chen:2019bwt, Dai:2020wny, Zhou:2022eft}. The nature of black hole chaos needs to be further explored.

Maldacena, Shenker and Stanford derived a universal temperature-dependent upper bound of the Lyapunov exponent $\lambda$ in quantum chaotic systems by the quantum field theory \cite{Maldacena:2015waa}
\be
\lambda \leq \frac{2 \pi T}{\hbar},
\ee
where $T$ is the temperature of the system. Such temperature-dependent characteristics can also be obtained through the thought experiment of the shock wave near the horizon of a black hole \cite{Shenker:2013pqa, Shenker:2013yza}, and some calculations of the shock wave have studied the Lyapunov exponent near the horizon \cite{Poojary:2018esz, Jahnke:2019gxr, Liu:2020yaf}. In  black hole calculations, the equivalent form of the chaos bound can be obtained by the natural unit $\hbar=1$ and the relationship between the Hawking temperature $T_H$ at the black hole's event horizon and the surface gravity $\kappa$
\be\label{cb}
\lambda \leq \kappa.
\ee
In the background of black holes, this upper bound can be tested in single-particle systems. Hashimoto and Tanahashi obtained a consistent upper bound of the Lyapunov exponent by considering the test particles maintain the static equilibrium due to external forces outside the black hole \cite{Hashimoto:2016dfz}. This result inspires the study of the Lyapunov exponent's upper bound outside black holes by considering particle motion.

In \cite{Zhao:2018wkl}, Zhao et al. studied the static equilibrium of charged particles near a large class of charged black holes and discussed the near-horizon expansion. They found that in the static equilibrium of the test particle, the bound \meqref{cb} is satisfied by Reissner–Nordstr\"om (RN) and Reissner–Nordstr\"om anti-de Sitter (RN-AdS) black holes and can be violated by some black holes \cite{Zhao:2018wkl}. The violation of the upper bound for the Lyapunov exponent was also found in the black hole with quasi-topological electromagnetism \cite{Lei:2020clg}. Taking into account the effects of angular momentum,  the circular motion of the test particle can violate the chaos bound for the RN black hole \cite{Lei:2021koj}, the Kerr-Newman black hole \cite{Kan:2021blg} and the Kerr-Newman AdS black hole \cite{Gwak:2022xje}.  Different black holes were discussed to investigate the violation of chaos bound \cite{Yu:2022tlr,Gao:2022ybw,Chen:2022tbb,Yin:2022mjv,Song:2022lhf,Chen:2023wph}. In \cite{Giataganas:2021ghs}, the Lyapunov exponent of particle motion near the black hole and cosmological horizons was investigated, and the author pointed out that the null energy conditions do not guarantee the satisfaction of chaos bounds.
 One of the interesting questions is what properties of black holes are associated with the Lyapunov exponent, which inspires us to study the Lyapunov exponent of particle motion.

In this paper, we investigate the relationship between the Lyapunov exponent and the chaos bound when the test particle maintains a stationary equilibrium in the radial direction near the charged black brane. We focus on the equilibrium in the radial direction of test particles. The charged black brane has the hyperscaling violating factor \cite{Alishahiha:2012qu}. The hyperscaling violating factor can bring interesting properties to space-time geometry. Other several hyperscaling violating black brane solutions were proposed \cite{Ge:2016sel, Ge:2016lyn,Ge:2017fix,Ge:2019fnj}. The influence of the lateral momentum is considered to explore the bound violation. Charged particles can maintain an equilibrium in the radial direction near the horizon by the repulsive force from the electric charge and the lateral momentum. We do not consider the backaction of particle motion on the background spacetime. The influence of the two characteristic parameters of the charged black brane, the dynamical exponent $z$ and the hyperscaling violating exponent $\theta$, on the Lyapunov exponent is discussed. We found that the chaos bound can be violated in the parameter range where $\theta$ and $z$ are small. In our previous work \cite{Lei:2021koj}, it was pointed out that the chaos bound can be violated in the case of near-extremal charged black holes, so here we discuss the effect of the temperature of the charged black brane on the Lyapunov exponent of particle motion. We also consider the effect of the null energy condition (NEC) on the study of the chaos, and the expression of NEC is $T_{\mu\nu}\xi^\mu\xi^\nu \geq 0$. Our numerical results show that the bound can be violated when the NEC is violated. In the extremal cases of zero-temperature and satisfying NEC, we derive the critical parameters $\theta_c$ and $z_c$, and find the chaos bound $\lambda \leq \kappa$ is satisfied when $\theta > \theta_c$ or $z>z_c$.

The rest of this paper is organized as follows. In section \ref{Lbb}, we review the background of the charged black branes with the hyperscaling violating factor and its parameter relationship of the temperature and NEC. In section \ref{Ly}, we derive the Lyapunov exponent $\lambda$ for test particles that maintain equilibrium in the radial direction using the Jacobian matrix. In section \ref{Nresult}, the numerical results of $\kappa^2-\lambda^2$ are shown. We discuss the influences of temperature $T_H$, parameters ($\theta$ and $z$) and the null energy condition to the Lyapunov exponent. The violation of the chaos bound is found. In section \ref{Cparameter}, the critical parameters $\theta_c$ and $z_c$ are obtained from the extremal cases of black brane temperature and satisfying NEC. We summarize the main conclusions in section \ref{conclusion}. In appendix \ref{VNEC}, we discuss the case of the black brane with the violation of NEC.

\section{Review of the charged black branes with the hyperscaling violating factor}\label{Lbb}
The charged black brane with the hyperscaling violating factor considered is a solution to the Einstein-Maxwell-Dilaton theory. The Einstein-Maxwell-Dilaton gravity has a $d+2$-dimensional minimal model with action \cite{Alishahiha:2012qu, Ge:2016sel,Ge:2019fnj}
\be
\label{action}
S=-\frac{1}{16\pi G}\int d^{d+2}x\sqrt{-g}\left[R-\frac{1}{2}(\partial\phi)^2+V(\phi)-\frac{1}{4}
\sum_{i=1}^2 e^{\lambda_i\phi}F_i^2\right],
\ee
where $R$ is the Ricci scalar, $\phi$ is the scalar field and $V(\phi)$ is the potential function of scalar field. The model has two free parameters $\lambda_1$, $\lambda_2$ and two gauge fields $F_1$, $F_2$.

The charged black brane solution with hyperscaling violating factor from the action \meqref{action} can be written as\cite{Alishahiha:2012qu}
\bea\label{metric}
ds^2&=&r^{-2\frac{\theta}{d}}\left(-r^{2z}f(r)dt^2+\frac{dr^2}{r^2f(r)}+r^2d\vec{x}^2\right),\cr &&\cr
 F_{1\;rt}&=&\sqrt{2(z-1)(z+d-\theta)}e^{\frac{\theta(1-d)/d+d}{\sqrt{2(d-\theta)(z-1-\theta/d)}}\phi_0}
\;r^{d+z-\theta-1},\cr &&\cr
F_{2\;rt}&=&Q\sqrt{2(d-\theta)(z-\theta+d-2)} e^{-\sqrt{\frac{z-1-\theta/d}{2(d-\theta)}}\;\phi_0}\;r^{-(z+d-\theta-1)},\cr &&\cr
e^{\phi}&=&e^{\phi_0}r^{\sqrt{2(d-\theta)(z-1-\theta/d)}},
\eea
where $\phi_0$ is a constant scalar field. The parameters $z$ and $\theta$ are the dynamical and hyperscaling violating exponents, respectively, with $z>1$ and $0 \leq \theta <d$. It should be noted that this solution will not be valid when $\theta=d$. The blacken factor $f(r)$ is given by
\be
f(r)=1-\frac{M}{r^{z+d-\theta}}+\frac{Q^2}{r^{2(z+d-\theta-1)}},
\ee
where $M$ is the mass of black brane and $Q$ is the electric charge. $F_{1\;rt}$ is an auxiliary gauge field and $F_{2\;rt}$ is the electric field. Taking infinity as the reference point of the electric potential, the electric potential function $A_t$ is given by
\be
A_t=\int_\infty^r F_{2\;rt} dr=-\frac{\sqrt{2}Qe^{-\phi_0\sqrt{\frac{d-dz+\theta}{2d(d-\theta)}}} r^{2-d-z+\theta}}{\sqrt{d+z-\theta-2}}
\ee
The constraint $d+z-\theta-2 \geq 0$ can ensure that the electric potential function is real. The radius of horizon $r_h$ can be defined by $f(r_h)=0$, which can lead to
\be
r_h^{2(d+z-\theta-1)}-M r_{h}^{d+z-\theta-2}+Q^2=0.
\ee
The Hawking temperature $T_H$ at the horizon is given by
\be\label{te}
T_H=\frac{(d+z-\theta)r_h^z}{4\pi}\left(1-\frac{(d+z-\theta-2)Q^2}{d+z-\theta}r_h^{2(\theta-d-z+1)}\right).
\ee
We can obtain the corresponding surface gravity $\kappa$, that is to say
\be
\kappa=2\pi T_H=\frac{(d+z-\theta)r_h^z}{2}\left(1-\frac{(d+z-\theta-2)Q^2}{d+z-\theta}r_h^{2(\theta-d-z+1)}\right).
\ee
To avoid naked singularity, the following inequality must be satisfied
\be\label{Tcon}
r_h^{2(d+z-\theta-1)}\geq\frac{(d+z-\theta-2)}{d+z-\theta}Q^2.
\ee

With the null vector $\xi^\mu=(\sqrt{g^{rr}},\sqrt{g^{tt}},0)$, the null energy condition (NEC) of the black brane \meqref{metric} is \cite{Alishahiha:2012qu}
\be
T_{\mu\nu}\xi^\mu\xi^\nu\sim d(\alpha+1)(\alpha+z-1)r^{-2\alpha} f(r) \geq 0,
\ee
where $\alpha=-\frac{\theta}{d}$. The null energy condition (NEC) can be recast as
\be\label{necon}
(\alpha+1)(\alpha+z-1)\geq 0.
\ee
Here we can obtain all the parameter conditions satisfied by the charged black brane with the hyperscaling violating factor
\be
\begin{aligned}
&d+z-\theta-2 \geq 0,\\
&r_h^{2(d+z-\theta-1)}\geq\frac{(d+z-\theta-2)}{d+z-\theta}Q^2, \\
&(\alpha+1)(\alpha+z-1)\geq 0 .
\end{aligned}
\ee
Among the parametric constraints, the null energy condition(NEC) is an important element, and it is one of the key parts of our discussion. Since the violation of NEC is controversial, we put the case of the spacetime with the violation of NEC in the appendix \ref{VNEC} as a referenceable supplementary discussion. In such background, the null energy condition $T_{\mu\nu}\xi^\mu\xi^\nu \geq 0$ is violated.

\begin{figure}[htbp]
    \centering
    \includegraphics[width=0.80\textwidth]{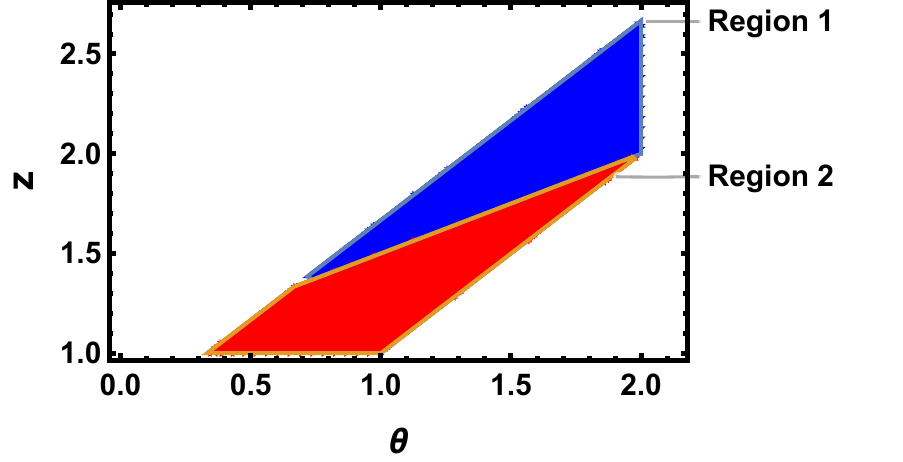}
    \caption{The available physical space $(\theta , z)$, when $d=2$, $\phi_0=0$, $Q=2$ and $r_h=1$. \textbf{Region 1} describes the charged black brane with hyperscaling violating factor. \textbf{Region 2} corresponds to the case where the null energy condition is violated.}
    \label{regionp}
\end{figure}

We focus on the effect of the hyperscaling violating exponent $\theta$ and the dynamical exponent $z$, so to simplify the calculation we set the parameters
\be
\phi_0=0 \qquad \text{and} \qquad r_h=1.
\ee
The inequality \meqref{Tcon} can be recast as
\be
1\geq \frac{(d+z-\theta-2)}{d+z-\theta}Q^2,
\ee
and there is no extremal black brane when $Q^2<1$. In the study of RN black holes, we pointed out the violation of chaos bound in the near-extremal RN black hole \cite{Lei:2021koj}. Thus it is interesting to discuss the Lyapunov exponent of particle motion in the extremal and near-extremal charged black brane. To explore the cases of extremal and near-extremal black brane and present results more clearly, we consider the charge of the black brane $Q=2$ in this paper.

In this work, we focus on the 4-dimensional cases, and we set the dimensional parameter $d=2$. The physical parameter space $(\theta , z)$ describing the charged black branes and the spacetime with the violation of NEC is plotted in \mpref{regionp}. As shown in the plot, \textbf{Region 1} in blue represents the charged black brane with the hyperscaling violating factor. Red \textbf{region 2} indicates the background where NEC is violated, which is discussed in appendix \ref{VNEC}. In the parameter space $(\theta , z)$, the bottom of charged black branes (\textbf{Region 1}) means $T_{\mu\nu}\xi^\mu\xi^\nu=0$, and under this boundary, NEC is broken.

\section{The Lyapunov exponent of charged particles with equilibrium in the radial direction}\label{Ly}
We focus on the equilibrium in the radial direction of test particles near the horizon, which means its radial position is a constant. Near a black hole, the equilibrium in the radial direction can be represented as the circular motion of test particles on the equatorial plane of the black hole. Static equilibrium is a special case of the equilibrium in the radial direction that indicates that test particles remain stationary in space. Near the horizon, when a particle in equilibrium is perturbed, its perturbation grows exponentially with time, and its corresponding exponent is the Lyapunov exponent \cite{Hashimoto:2016dfz}. From the effective potential analysis, this instability of the test particles' equilibrium in the radial direction corresponds to a local maximum of the effective potential of test particles. More discussion of the effective potential about particle motion near horizon can be found in \cite{Kan:2021blg,Gwak:2022xje,Yu:2022tlr}.

The Lyapunov exponent of test particle also can be calculated by using the Jacobian matrix \cite{Cardoso:2008bp, Pradhan:2012rkk, Pradhan:2013bli}. Here we calculate the Lyapunov exponent of test particles that maintain an equilibrium in the radial direction near the charged black brane. For simplicity, we rewrite the black brane's metric \meqref{metric} as
\be\label{smetric}
ds^2=-F(r)dt^2+\frac{dr^2}{H(r)}+G(r)d\vec{x}^2,
\ee
where $F(r)=r^{2\left(z-\frac{\theta}{d} \right)}f(r)$, $H(r)=r^{2\left(1+\frac{\theta}{d} \right)}f(r)$ and $G(r)=r^{2\left(1-\frac{\theta}{d} \right)}$. When we focus on the 4-dimensional case, which can result in $d=2$ and $d\vec{x}^2=dx^2+dy^2$. Consider a charged particle moving in the $y=0$ plane, its Lagrangian can be written
\be
\mathcal{L}=\frac{1}{2}\left(-F(r)\dot{t}^2+\frac{\dot{r}^2}{H(r)}+G(r)\dot{x}^2 \right)- q A_t (r) \dot{t},
\ee\label{cL}
where the dot denotes a derivative with respect to the proper time $\tau$. The generalized momenta $\pi_\mu=\frac{\partial \mathcal{L}}{\partial \dot{x}^\mu}$ are
\be
\begin{aligned}
\pi_t=&-(F(r)\dot{t}+qA_t(r))=-E=\text{Constant},
\\
\pi_r=&\frac{\dot{r}}{H(r)},
\\
\pi_x=&G(r)\dot{x}=\text{Constant},
\end{aligned}
\ee
where $E$ is the energy of the test particle, $\pi_r$ is the radial momentum and $\pi_x$ is the lateral momentum. The different values of constants $(q,\ \pi_t,\ \pi_x)$ of test particles result in different states of particle motion, such as falling into the horizon, moving away from the horizon, periodic motion, and stationary equilibrium, etc.

With the formula $\mathcal{H}=\pi_\mu \dot{x}^\mu-\mathcal{L}$, the Hamiltonian of the test particle is
\be
\mathcal{H}=\frac{1}{2}\left(-\frac{(\pi_t+qA_t(r))^2}{F(r)}+H(r)\pi_r^2+\frac{\pi_x^2}{G(r)} \right),\label{cH}
\ee
which  leads to the canonical equations of motion for the test particle
\be
\dot{x}^\mu=\frac{\partial \mathcal{H}}{\partial \pi_\mu},\qquad \dot{\pi}_\mu=-\frac{\partial \mathcal{H}}{\partial x^\mu}.
\ee
The radial evolution equations of the test particle with respect to the coordinate time $t$ are
\begin{small}
\be
\begin{aligned}
\frac{dr}{dt}=&\frac{\dot{r}}{\dot{t}}=-\frac{\pi_r F(r) H(r)}{\pi_t+q A_t(r)},
\\
\frac{d\pi_r}{dt}=&\frac{\dot{\pi}_r}{\dot{t}}=\frac{1}{2}\left(\frac{(\pi_t +q A_t(r))F(r)^{'}}{F(r)}+\frac{F(r)(H(r)^{'}G(r)^2\pi_r^2-G(r)^{'}\pi_x^2)}{(\pi_t+qA_t(r))G(r)^2} -2q A_t(r)^{'}\right),
\end{aligned}\label{req}
\ee
\end{small}
where the prime `` $\prime$ '' denotes derivative with respect to r. We can consider the four-velocity normalization condition
\be
g_{\mu \nu }\dot{x}^\mu \dot{x}^\nu=\eta,
\ee
where $\eta$ is the normalization constant with $\eta=-1$ the time-like orbits, $\eta=0$ the null orbits and $\eta=1$ the space-like orbits. Here, we consider the charged test particle moving along the time-like orbits and the null orbits.

We can obtain the Jacobian matrix of test particle motion by taking ($r$, $\pi_{r}$) as the phase space variables. For the convenience of writing, we will mark the equations \meqref{req} as $\frac{d r}{dt}=M_1$ and $\frac{d \pi_{r}}{dt}=M_2$. The Jacobian matrix $K_{ij}$ can be defined by
\be
K_{ij}=\left(
\begin{aligned}
\frac{\partial M_1}{\partial r} \quad &\frac{\partial M_1}{\partial \pi_{r}}
\\
\frac{\partial M_2}{\partial r} \quad &\frac{\partial M_2}{\partial \pi_{r}}
\end{aligned}
\right).
\ee
For the equilibrium in the radial direction of test particles, it should satisfy the equilibrium condition $\frac{d r}{dt}=\frac{d \pi_r}{dt}=0$. The Jacobian matrix $K_{ij}$ of test particles can be reduced at the equilibrium position $r=r_0$. The components are
\begin{small}
\be
\begin{aligned}
K_{11}=&0,
\\
K_{12}=&\left.-\frac{F(r) H(r)}{\pi_t +q A_t(r)}\right|_{r=r_0},
\\
K_{21}=&\left.-\frac{1}{2}\left(2 q A_t(r)^{''}-\left(\frac{(\pi_t+q A_t(r))F(r)^{'}}{F(r)} \right)^{'}+\left(\frac{\pi_x^2 F(r) G(r)^{'}}{(\pi_t +q A_t(r))G(r)^{2}}\right)^{'} \right)\right|_{r=r_0},
\\
K_{22}=&0.
\end{aligned}
\ee
\end{small}
The eigenvalues of the Jacobian matrix $K_{ij}$ can lead to the Lyapunov exponent $\lambda$ at the equilibrium position $r=r_0$
\begin{footnotesize}
\be\label{Lye}
\lambda^2=\left.\frac{F(r) H(r)}{2(\pi_t +q A_t(r))}\left(2 q A_t(r)^{''}-\left(\frac{(\pi_t+q A_t(r))F(r)^{'}}{F(r)} \right)^{'}+\left(\frac{\pi_x^2 F(r) G(r)^{'}}{(\pi_t +q A_t(r))G(r)^{2}}\right)^{'} \right)\right|_{r=r_0}.
\ee
\end{footnotesize}
We discuss the effect of the lateral momentum $\pi_x$ on the particle motion, and the relationship between the constants $q$, $\pi_t$ and $\pi_x$ in the above equation is contracted by the equilibrium condition $\frac{d r}{dt}=\frac{d \pi_r}{dt}=0$. We can see that the test particle at the equilibrium position $r=r_0$ satisfies
\begin{small}
\be\label{cons}
\begin{aligned}
\pi_t=&\left.\frac{G(r)(\eta G(r)-\pi_x^2)(2F(r) A_t(r)^{'}-A_t(r) F(r)^{'})-\pi_x^2 A_t(r) F(r) G(r)^{'}}{2G(r)A_t(r)^{'}\sqrt{F(r)G(r)(\pi_x^2-\eta G(r))}}\right|_{r=r_0},
\\
q=&\left.\frac{G(r)F(r)^{'}(\eta G(r)-\pi_x^2)+\pi_x^2F(r)G(r)^{'}}{2G(r)A_t(r)^{'}\sqrt{F(r)G(r)(\pi_x^2-\eta G(r))}}\right|_{r=r_0}.
\end{aligned}
\ee
\end{small}
Taking the values of $\pi_t$ and $q$ obtained from the above equations into \meqref{Lye}, we can analyze the effect of $\pi_x$ in the Lyapunov exponent. The normalization constant $\eta =-1,\ 0$ corresponds to the time-like orbits and the null orbits, respectively. We label the Lyapunov exponent as $\lambda_s$ when the particle maintains a static equilibrium (the lateral momentum $\pi_x=0$), the Lyapunov exponent for time-like orbits as $\lambda_t$, and the Lyapunov exponent for null orbits as $\lambda_n$. From \meqref{Lye} and \meqref{cons}, we can obtain $\lim\limits_{\pi_x \rightarrow \infty } \lambda_t = \lambda_n$, which means the Lyapunov exponent of time-like orbits is equal to that of the null orbit when $\pi_x \rightarrow \infty$. So when we want to discuss the Lyapunov exponent of massive particles in the limit that $\pi_x \rightarrow \infty$, we can consider $\lambda_n$ of null orbit.

\section{The analysis of the Lyapunov exponent near charged black branes}\label{Nresult}
\begin{figure}[htb]
    \centering
    \includegraphics[width=0.80\textwidth]{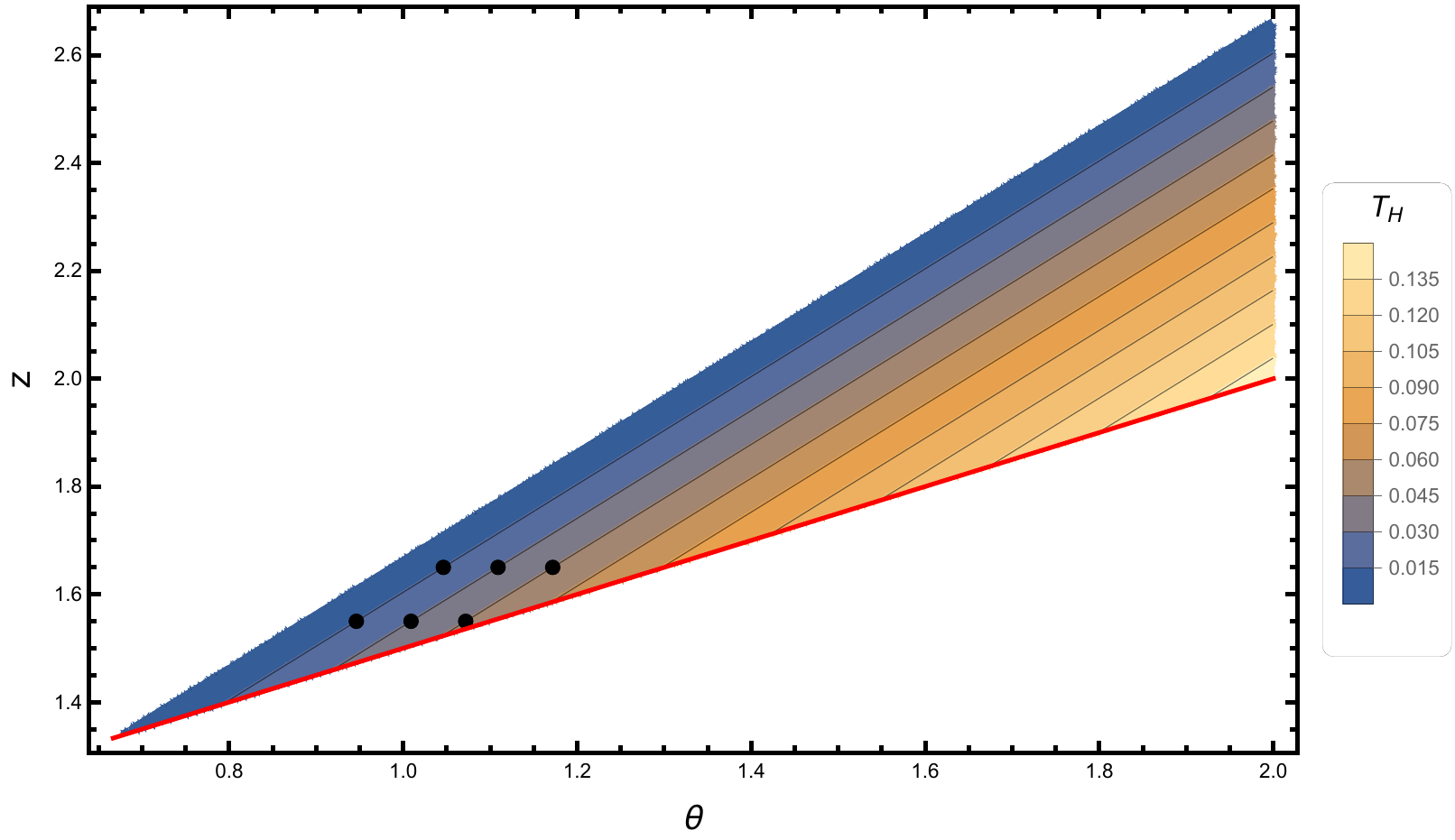}
    \caption{The temperature $T_H$ of the charged black brane as a function of the parameters $\theta$ and $z$. The black dots ($\theta$, $z$) in the figure are the parameter values for the black branes that we discuss next subsection.}
    \label{PT1}
\end{figure}

In this section, we discuss the relationship between the chaos bound and the Lyapunov exponent for charged particles which maintain equilibrium in the radial direction near charged black branes with the hyperscaling violating factor. We explore the effect of the hyperscaling violating exponent $\theta$ and the dynamical parameter $z$. The influence of the black brane temperature and NEC is also discussed. As shown in the previous section, we set $d=2,\qquad \phi_0=0,\qquad Q=2,\qquad r_h=1$. The corresponding valid parameter space $(\theta , z)$ with the temperature $T_H$ is shown in \mpref{PT1}.

As shown in \mpref{PT1}, when $\theta$ is a constant, the black brane temperature $T_H$ decreases as $z$ increases; when z is a constant, the black brane temperature $T_H$ increases as $\theta$ increases. The red line in the figure shows the cases of $T_{\mu\nu}\xi^\mu\xi^\nu=0$, and the NEC is violated below the red line. The points where $z$ is smaller or $\theta$ is larger are closer to the region where the null energy condition is violated. To express our results more clearly, we discuss whether the chaos bound is violated by evaluating $\kappa^2-\lambda^2$ numerically. When $\kappa^2-\lambda^2 < 0$, the chaos bound $\lambda \leq \kappa$ is violated. The Lyapunov exponent $\lambda$ is calculated by \meqref{Lye}.

\subsection{Fixed temperature $T_H$, varying $\theta$, $z$ }
With fixed temperature $T_H$, but varying $\theta$, $z$ for the charged black branes, we first numerically analyze the relationship between the chaos bound and the Lyapunov exponent corresponding to the equilibrium in the radial direction. We consider the temperature of the black brane $T_H=0.015,\ 0.030,\ 0.045$. The parameter values in the parameter space $(\theta ,z)$ are shown in \mpref{PT1} with the black dots.

\begin{figure}[htb]
\centering
\subfigure[$\theta =0.946$, $z=1.550$]{\label{R1T1a}
\begin{minipage}[t]{0.45\linewidth}
\centering
\includegraphics[width=1 \textwidth]{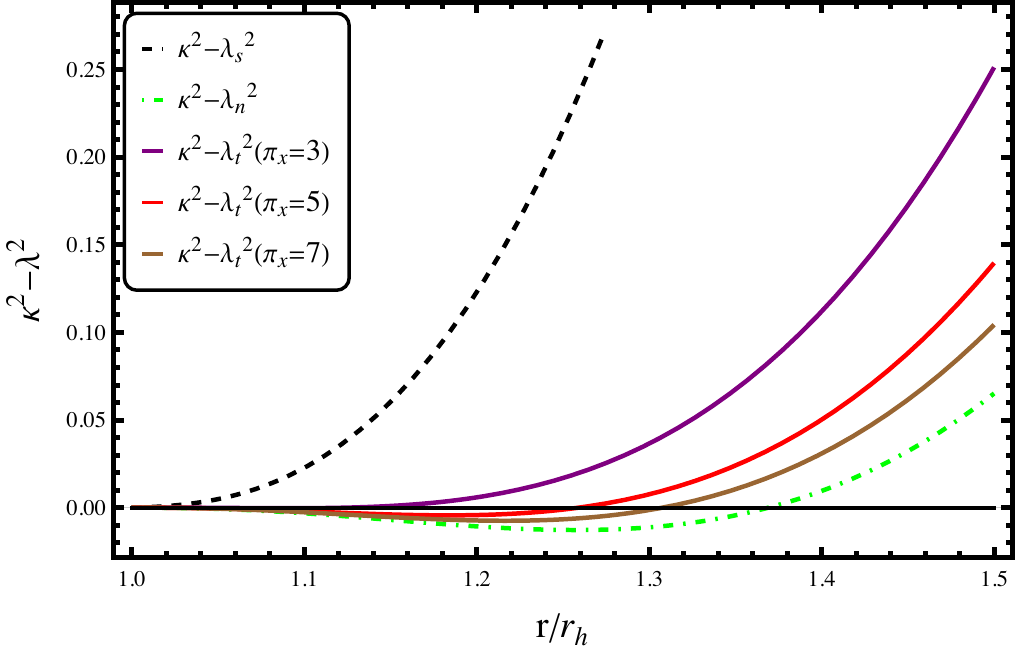}
\end{minipage}
}
\subfigure[$\theta=1.046$, $z=1.650$]{\label{R1T1b}
\begin{minipage}[t]{0.45\linewidth}
\centering
\includegraphics[width=1 \textwidth]{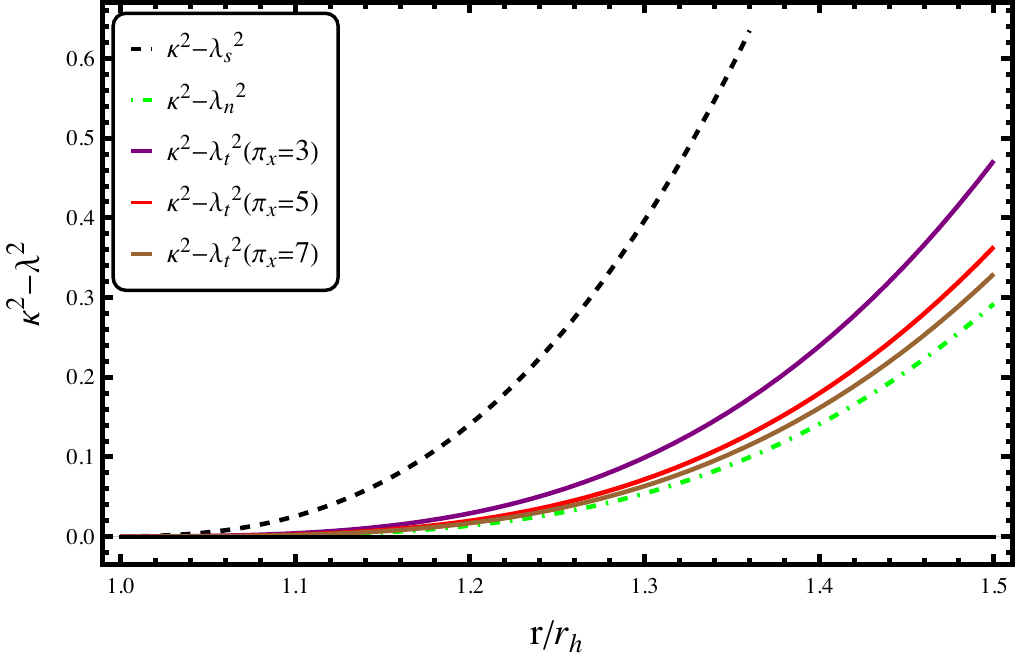}
\end{minipage}
}
\caption{ $\kappa^2-\lambda^2$ as a function of $r/r_h$ near the charged black brane with (a) $\theta =0.946$, $z=1.550$ and (b) $\theta=1.046$, $z=1.650$ at $T_H=0.015$. The chaos bound is violated in \mpref{R1T1a} since $\kappa^2-\lambda^2<0$.}
\label{R1T1}
\end{figure}

\begin{figure}[htb]
\centering
\subfigure[$\theta =1.009$, $z=1.550$]{\label{R1T2a}
\begin{minipage}[t]{0.45\linewidth}
\centering
\includegraphics[width=1 \textwidth]{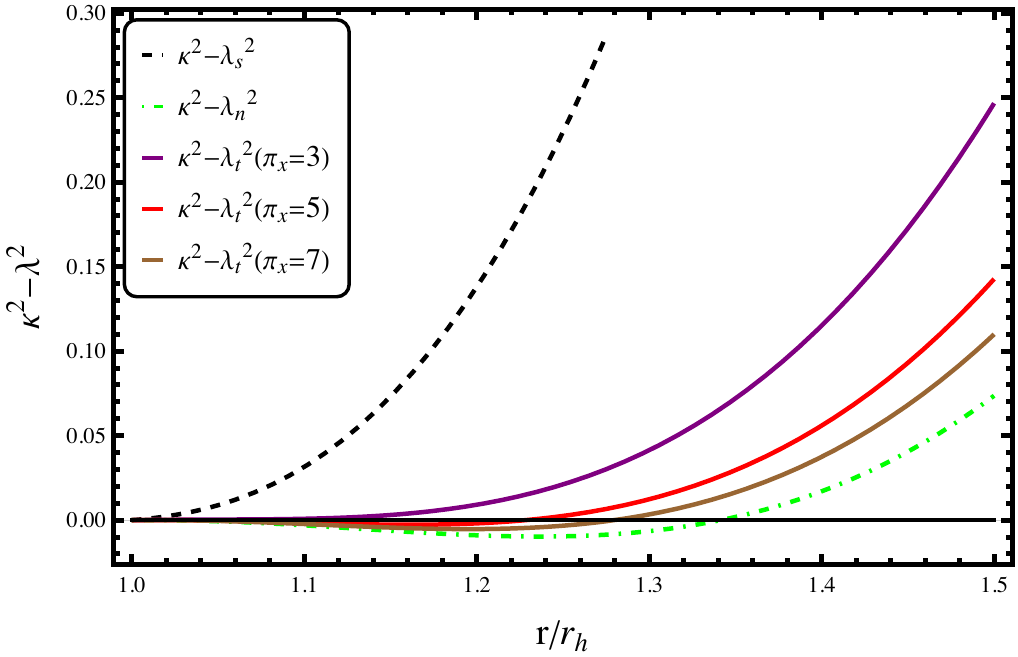}
\end{minipage}
}
\subfigure[$\theta =1.109$, $z=1.650$]{\label{R1T2b}
\begin{minipage}[t]{0.45\linewidth}
\centering
\includegraphics[width=1 \textwidth]{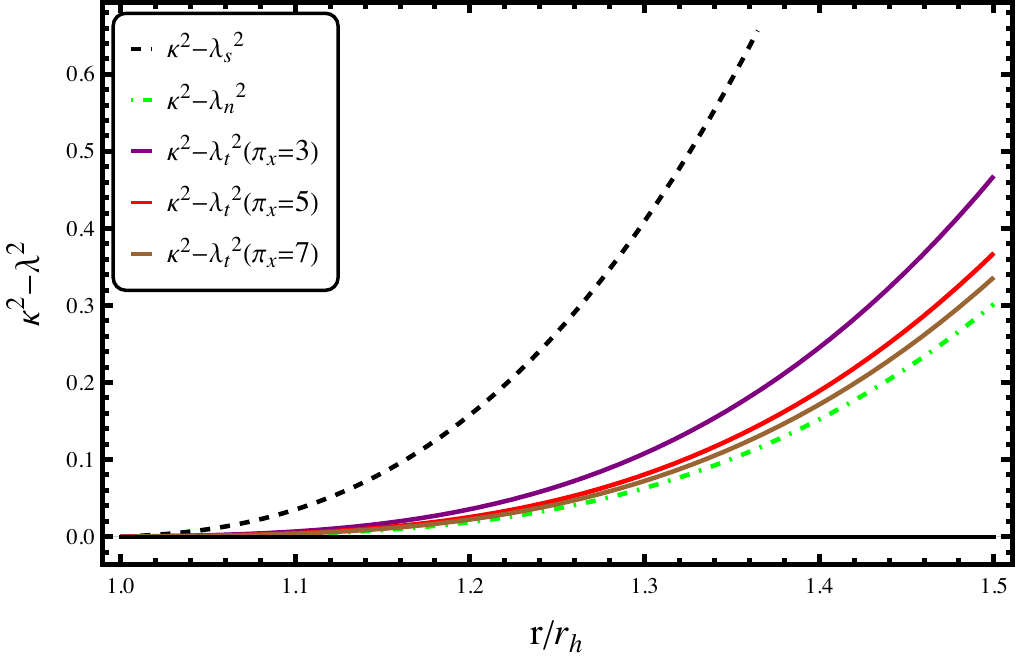}
\end{minipage}
}
\caption{$\kappa^2-\lambda^2$ as a function of $r/r_h$ near the charged black brane with (a) $\theta =1.009$, $z=1.550$ and (b) $\theta=1.109$, $z=1.650$ at $T_H=0.030$. The chaos bound is violated for the green line in \mpref{R1T2a} since $\kappa^2-\lambda^2<0$.}
\label{R1T2}
\end{figure}

\begin{figure}[htb]
\centering
\subfigure[$\theta =1.072$, $z=1.550$]{\label{R1T3a}
\begin{minipage}[t]{0.45\linewidth}
\centering
\includegraphics[width=1 \textwidth]{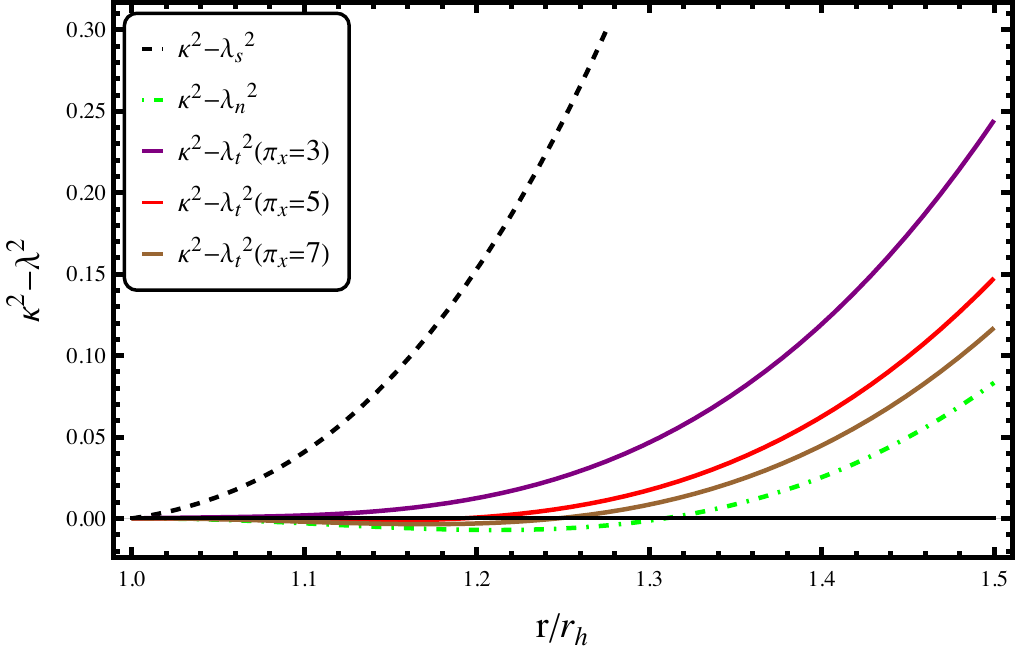}
\end{minipage}
}
\subfigure[$\theta =1.172$, $z=1.650$]{\label{R1T3b}
\begin{minipage}[t]{0.45\linewidth}
\centering
\includegraphics[width=1 \textwidth]{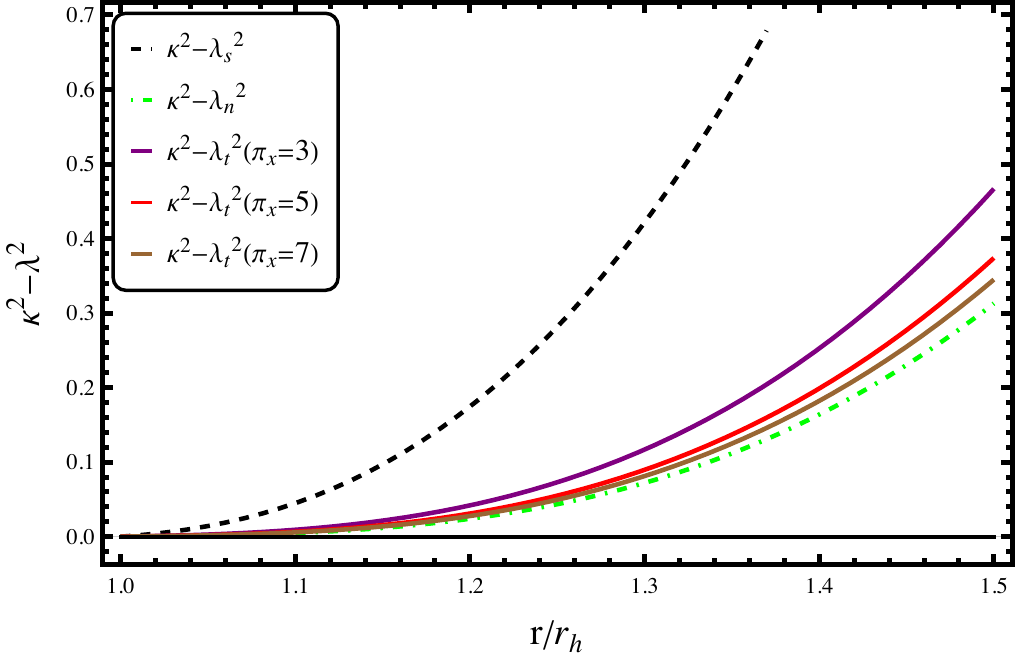}
\end{minipage}
}
\caption{$\kappa^2-\lambda^2$ as a function of $r/r_h$ near the charged black brane with (a) $\theta =1.072$, $z=1.550$ and (b) $\theta=1.172$, $z=1.650$ at $T_H=0.045$. The chaos bound is violated for the green line in \mpref{R1T3a} since $\kappa^2-\lambda^2<0$.}
\label{R1T3}
\end{figure}

For a massive particle, we consider its static equilibrium (the lateral momentum $\pi_x=0$) and the finite lateral momentum ($\pi_x=3,\ 5,\ 7$).
The Lyapunov exponent of null orbits is also considered, which equals to the value of time-like orbits with $\pi_x \rightarrow \infty$. The numerical results are presented in Fig. \ref{R1T1}-\ref{R1T3}. In these figures, we use dashed lines to show the static equilibrium of a massive particle, dot-dashed lines to show the null orbits, and solid lines to show the results of the massive particle with finite lateral momentum ($\pi_x=3,\ 5,\ 7$), respectively. $\kappa^2-\lambda^2<0$ in these figures indicates that the chaos bound is violated by the test particle with equilibrium in the radial direction.

In \mpref{R1T1}, we plot the numerical results of $\kappa^2-\lambda^2$ for the black brane temperature $T_H=0.015$. We can see from \mpref{R1T1a} that there is $\kappa^2-\lambda^2<0$ for $\lambda_n$ and $\lambda_t$, which indicates that the chaos bound is violated; while in \mpref{R1T1b} the chaos bound is not violated. This result indicates that the parameters $\theta$ and $z$ affect the equilibrium stability of the test particles at the same temperature. The numerical results of $\kappa^2-\lambda^2$ for the black brane temperature $T_H=0.030,\ 0.045$ are plotted in \mpref{R1T2} and \mpref{R1T3}, respectively. Similar to the case for $T_H=0.015$, the chaos bound can be violated at the same temperature when $\theta$ and $z$ are small. Meanwhile, comparing \mpref{R1T1a}, \mpref{R1T2a} and \mpref{R1T3a}, we can see that the higher temperature $T_H$ is, the smaller the range $r/r_h$ of the chaos bound violation is.

In Fig. \ref{R1T1}-\ref{R1T3}, at the same equilibrium position, $\kappa^2-\lambda^2$ decreases with the increase of the lateral momentum $\pi_x$ of the test particle, then it indicates that the Lyapunov exponent of the time-like orbits becomes stronger with the increase of $\pi_x$. The Lyapunov exponent of the null orbits is the largest, which is consistent with our previous results of RN black holes \cite{Lei:2021koj}. In the charged black brane background, the parameters of the black brane, in addition to the temperature, also affect whether the chaos bound is violated.

Next, we will further discuss the relationship between the black brane parameters and the Lyapunov exponent of the test particle with equilibrium in the radial direction. We discuss the null orbit because it has the largest Lyapunov exponent and is most likely to exceed the bound $\lambda \leq \kappa$.

\subsection{Fixed parameter $\theta$ or $z$}
\begin{figure}[htb]
\centering
\subfigure[$\theta=0.80$]{\label{R1tha}
\begin{minipage}[t]{0.31\linewidth}
\centering
\includegraphics[width=1 \textwidth]{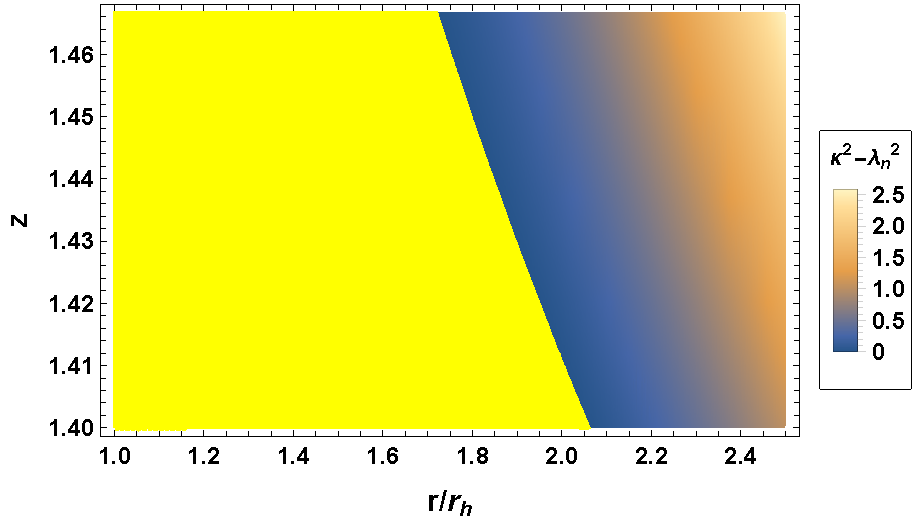}
\end{minipage}
}
\subfigure[$\theta=1.10$]{\label{R1thb}
\begin{minipage}[t]{0.31\linewidth}
\centering
\includegraphics[width=1 \textwidth]{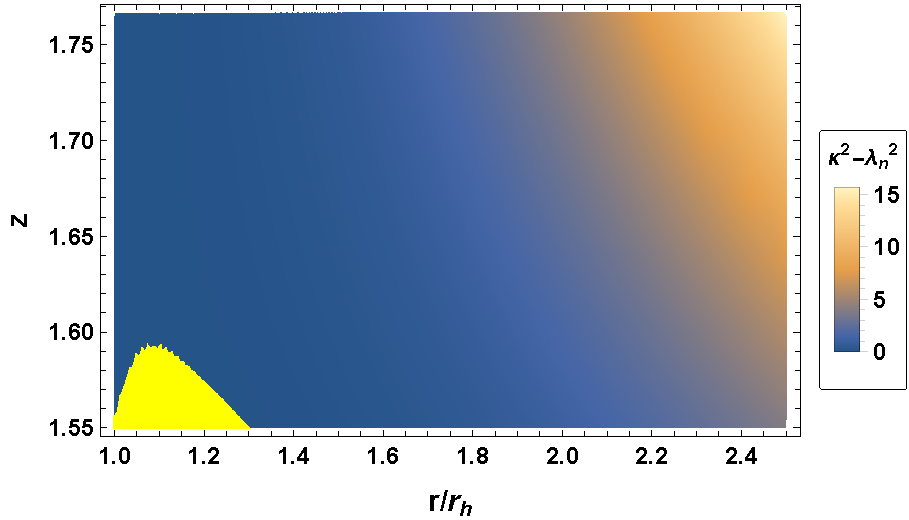}
\end{minipage}
}
\subfigure[$\theta=1.50$]{\label{R1thc}
\begin{minipage}[t]{0.31\linewidth}
\centering
\includegraphics[width=1 \textwidth]{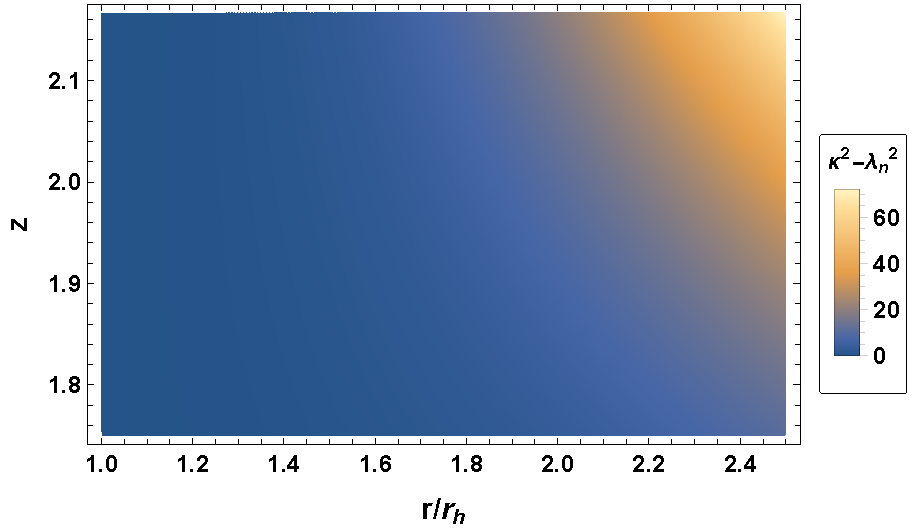}
\end{minipage}
}
\caption{The contour plot of $\kappa^2-\lambda_n^2$ as a function of $z$ and $r/r_{h}$ for fixed $\theta$: (a) $\theta=0.80$, (b) $\theta=1.10$ and (c) $\theta=1.50$. The yellow region corresponds to the chaos bound violated region.}
\label{R1th}
\end{figure}

\begin{figure}[htb]
\centering
\subfigure[$z=1.40$]{\label{R1za}
\begin{minipage}[t]{0.31\linewidth}
\centering
\includegraphics[width=1 \textwidth]{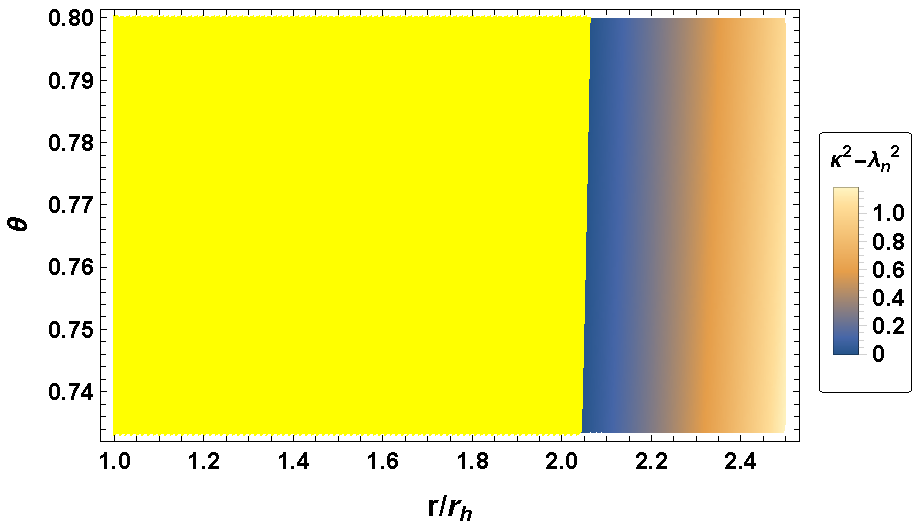}
\end{minipage}
}
\subfigure[$z=1.59$]{\label{R1zb}
\begin{minipage}[t]{0.31\linewidth}
\centering
\includegraphics[width=1 \textwidth]{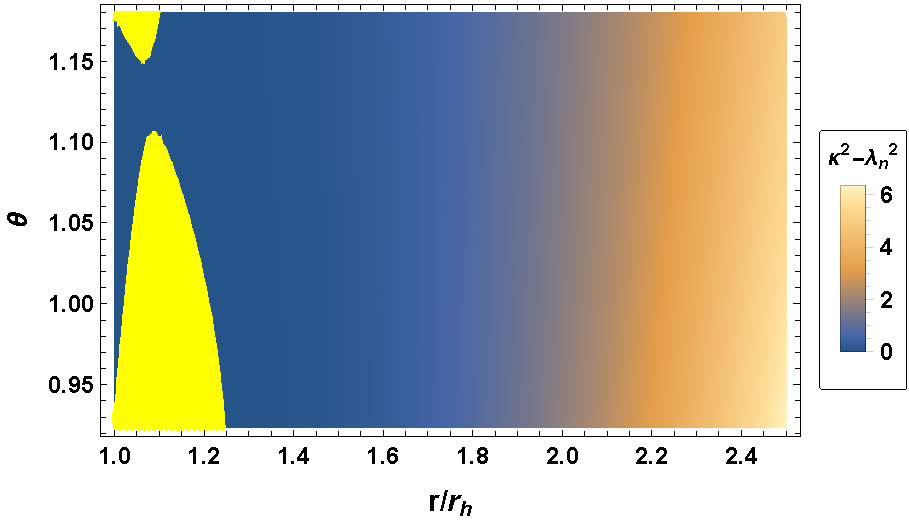}
\end{minipage}
}
\subfigure[$z=2.10$]{\label{R1zc}
\begin{minipage}[t]{0.31\linewidth}
\centering
\includegraphics[width=1 \textwidth]{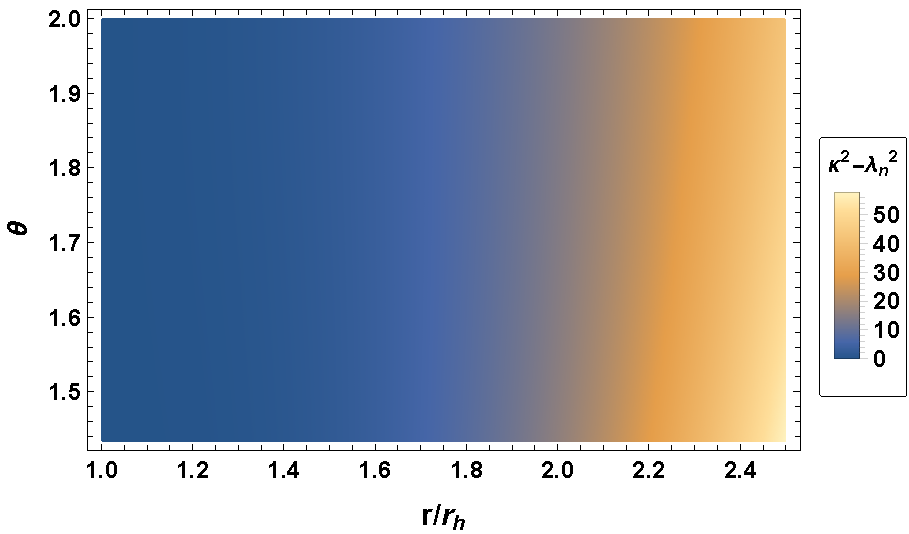}
\end{minipage}
}
\caption{The contour plot of $\kappa^2-\lambda_n^2$ as a function of $\theta$ and $r/r_h$ for fixed $z$: (a) $z=1.40$, (b) $z=1.59$ and (c) $z=2.10$.  The yellow region corresponds to the chaos bound violated region.}
\label{R1z}
\end{figure}

To explore the relationship between the black brane parameters and the Lyapunov exponent of particle motion, here we investigate the null orbits of the test particle at the same black brane parameter $\theta$ and $z$, respectively. The numerical results for $\kappa^2-\lambda_n^2$ are shown in \mpref{R1th} and \mpref{R1z}, with the colored region where $\kappa^2-\lambda_n^2>0$. The region where $\kappa^2-\lambda_n^2<0$ is marked in yellow in these plots, which indicates that the chaos bound is violated.

In \mpref{R1th}, the numerical results of $\kappa^2-\lambda_n^2$ for null orbits are shown, which has three plots for different $\theta=0.80,\ 1.10,\ 1.50$. For each plot, the horizontal axis represents $r/r_h$, and the vertical axis represents the parameter $z$. For the plot with $\theta =0.80$ in \mpref{R1tha}, all available values of $z$ have yellow regions, which means that the chaos bound is always violated. As $z$ increases, there is a decrease in the range of $r/r_h$ where the chaos bound can be violated. In \mpref{R1thb}, we show that for $\theta=1.10$, the chaos bound can be violated near the minimal value of $z$, and the region where the chaos bound is violated decreases as $z$ increases. The minimal value of $z$ corresponds to the zero-temperature of the charged black brane. There is no violation of the chaos bound as shown in \mpref{R1thc}.

For $z=1.40,\ 1.59,\ 2.10$, the results of $\kappa^2-\lambda^2$ are shown in \mpref{R1z}. The horizontal axis represents $r/r_h$, and the vertical axis represents the parameter $\theta$. We color the region in yellow where the chaos bound is violated. As shown in \mpref{R1za} with $z=1.40$, the chaos bound is violated for all values of $\theta$. In \mpref{R1zb}, there is $\theta=1.59$. The chaos bound can be violated in two regions close to the maximal and minimal values of $\theta$. The maximal and minimal values of $\theta$ correspond to the cases of $T_{\mu\nu}\xi^\mu\xi^\nu=0$ and the zero-temperature, respectively. The range $r/r_h$ of the violation of chaos bound is larger near the maximal and minimal values of $\theta$. The chaos bound is not violated for the plot with $\theta=2.10$ in \mpref{R1zc}.

From \mpref{R1th} and \mpref{R1z}, we can conclude these results. There is always $\kappa^2-\lambda_n^2 <0$ when $\theta$ and $z$ are small, as in \mpref{R1tha} and \mpref{R1za}. We can see that in \mpref{R1thb}\footnote{In this plot, there is no $\kappa^2-\lambda_n^2<0$ near the zero-temperature. This is probably because $z$ is already greater than some critical value. In the next discussion, there will be more intuitive results to show that $\lambda_n$ is more likely to exceed $\lambda \leq \kappa$ near extremal black branes.} and \mpref{R1zb}, as $\theta$ and $z$ increase, the violation of chaos bound only exists in the parameter range near the zero-temperature and the critical values of NEC. The range of $r/r_h$ with $\kappa^2-\lambda_n^2<0$ is the biggest at the extremal cases of temperature and NEC. As $\theta$ and $z$ deviate from the extremal cases, the range $r/r_h$ of $\kappa^2-\lambda_n^2<0$ decreases. In \mpref{R1thb} and \mpref{R1zb}, around the equilibrium position $r/r_h=1.1$, there is a maximal range of parameters that can violate $\lambda \leq \kappa$. In \mpref{R1thc} and \mpref{R1zc}, $\kappa^2-\lambda_n^2$ is always positive. There seems to be no more $\kappa^2-\lambda_n^2<0$ once $\theta$ and $z$ are large enough.

From these results, we can see that the violation of the chaos bound is influenced by the parameters $\theta$ and $z$ near the charged black brane, and the chaos bound is easily violated when $\theta$ and $z$ are small. Also, we can see that the chaos bound is also easily violated near the zero-temperature and the critical values of the null energy condition, which reveals the relationship between the particle motion instability and the black brane's parameters, temperature and the null energy condition.

\subsection{Fixed equilibrium position $r/r_h$}
To show more clearly the effect of the temperature and the null energy condition on the chaos bound, we analyze $\kappa^2-\lambda_n^2$ for the fixed radial position $r/r_h$ where the test particle maintains equilibrium in the radial direction.

\begin{figure}[!htbp]
\centering
\subfigure[$r/r_h=1.005$]{\label{R1ra}
\begin{minipage}[t]{0.48\linewidth}
\centering
\includegraphics[width=1 \textwidth]{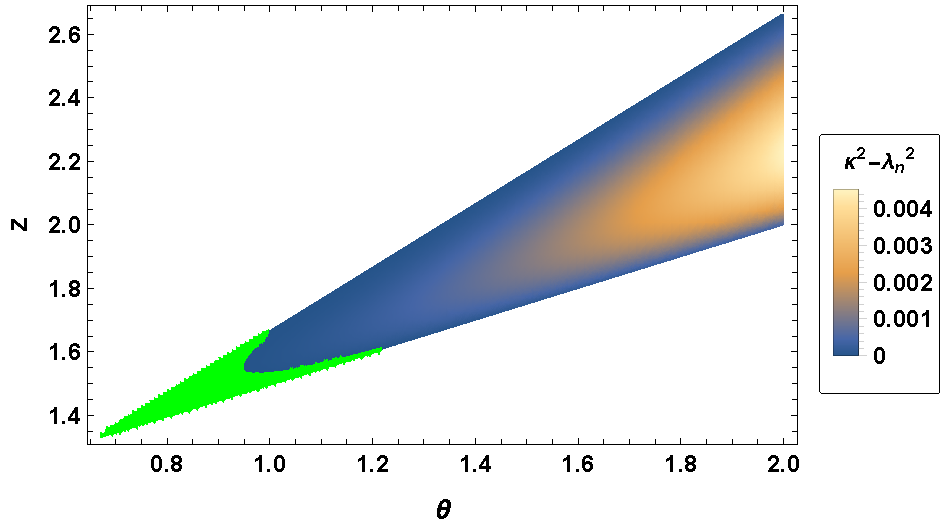}
\end{minipage}
}
\subfigure[$r/r_h=1.1$]{\label{R1rb}
\begin{minipage}[t]{0.48\linewidth}
\centering
\includegraphics[width=1\textwidth]{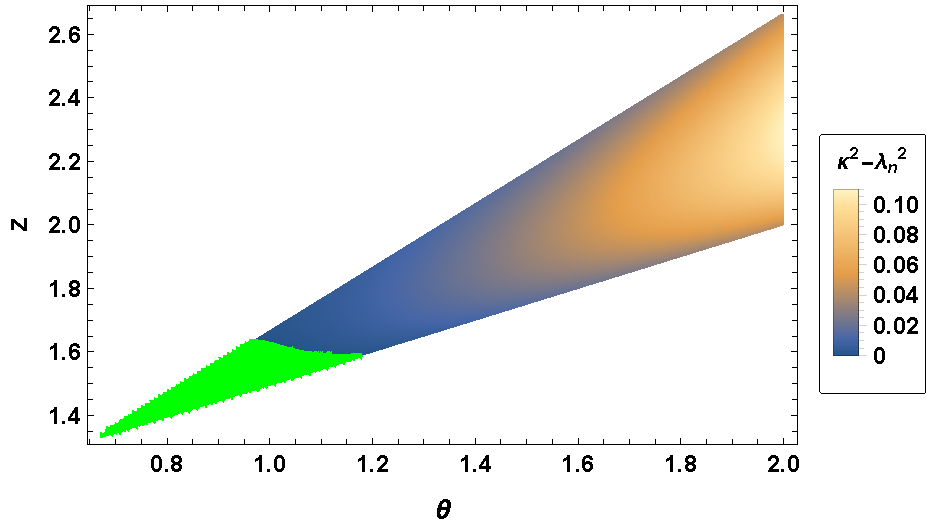}
\end{minipage}
}
\caption{The contour plot of $\kappa^2-\lambda_n^2$  as a function of $\theta$ and $z$ for fixed $r/r_h$: (a) $r/r_h=1.005$, (b) $r/r_h=1.1$. The green region cprresponds to the parameter space $(\theta , z)$ where the chaos bound is violated.}
\label{R1r}
\end{figure}

In \mpref{R1r}, the numerical result of $\kappa^2-\lambda_n^2$ as a function of $\theta$ and $z$ for fixed $r/r_h$ is plotted. The case of $r/r_h=1.005$ is considered because it is near the horizon. In the previous subsection, we can see that the violation of the chaos bound is often located at $r/r_h=1.1$, so we also consider the case of $r/r_h=1.1$. The region where $\kappa^2-\lambda_n^2<0$ is colored in green, which means the bound $\lambda \leq \kappa$ is violated.

The numerical results at the equilibrium position $r/r_h=1.005$ are shown in \mpref{R1ra}. There is always $\kappa^2-\lambda_n^2<0$ in the region where the parameters $\theta$ and $z$ are small. However, as the parameters increase, $\kappa^2-\lambda_n^2<0$ can be seen only in the parameter space near the zero-temperature and the cases of $T_{\mu\nu}\xi^\mu\xi^\nu=0$. The maximal value of $\theta=1.217$ for the violation of chaos bound at $r/r_h=1.005$ is given by $T_{\mu\nu}\xi^\mu\xi^\nu=0$, and another parameter maximal value $z=1.665$ is given by the zero-temperature limit. In \mpref{R1rb}, the result at the position $r/r_h=1.1$ shows that $\kappa^2-\lambda_n^2<0$ exists in the small parameter region $(\theta , z)$. Something is different in that there is still $\kappa-\lambda_n^2<0$ even in the middle parameter region. In this case, the maximal values of the parameters satisfying $\kappa^2-\lambda_n^2<0$ are $\theta=1.179$ and $z=1.634$, which are smaller than \mpref{R1ra}.

In \mpref{R1r} we can see that the value of $\kappa-\lambda_n^2$ is always smaller near the upper and lower boundary of the parameter space. The maximal parameter values that allow $\kappa-\lambda_n^2<0$ also always appear on the upper and lower boundary of the parameter space. The upper and lower boundary corresponds to the cases of the zero-temperature $T_H=0$ and $T_{\mu\nu}\xi^\mu\xi^\nu=0$, respectively. This suggests a connection between the Lyapunov exponent in the particle motion and the black brane temperature as well as the NEC. We can obtain the critical parameters $\theta_c$ and $z_c$ for $\kappa-\lambda_n^2<0$ in the cases of zero-temperature and critical values of NEC. These parameters $\theta_c$ and $z_c$ provide the possibility of the violation of chaos bound, i.e., the bound $\lambda \leq \kappa$ cannot be violated by test particles with equilibrium in the radial direction when $\theta>\theta_c$ and $z>z_c$.

\section{The critical parameters $\theta_c$ and $z_c$ for the violation of $\lambda \leq \kappa$ }\label{Cparameter}

In the previous section, we discussed the bound of test particles' equilibrium in the radial direction and found the violation. The effect of the black brane temperature $T_H$ and the null energy condition (NEC) is investigated. The results showed that the bound $\lambda \leq \kappa$ is more likely to be violated near the cases of zero-temperature black branes and the critical value of NEC ($T_{\mu\nu}\xi^\mu\xi^\nu=0$)\footnote{The null energy condition (NEC) means $T_{\mu\nu}\xi^\mu\xi^\nu \geq 0$, so we call the case of $T_{\mu\nu}\xi^\mu\xi^\nu=0$ as the critical value for the violation of NEC.}. Through the investigation of zero-temperature cases and the cases of $T_{\mu\nu}\xi^\mu\xi^\nu=0$, we can understand more about the violation of chaos bound in the charged black brane with the hyperscaling violating factor and find the critical parameters ($\theta_c$ and $z_c$) for the violation of chaos bound. When $\theta >\theta_c$ or $z >z_c$, the bound $\lambda \leq \kappa$ is always satisfied.

As before, we set the parameters $d=2$, $\phi_0=0$, $Q=2$ and $r_h=1$ in \meqref{metric}. We investigate whether the equilibrium in the radial direction of test particles violates the upper bound of the Lyapunov exponent $\lambda$ based on the value of $\kappa^2-\lambda^2$, and the bound is violated when $\kappa^2-\lambda^2<0$.

\subsection{$T_H=0$ case }
We consider the extremal black brane with $T_H=0$ and explore the violation of bound in test particles' equilibrium in the radial direction. From \meqref{Tcon}, we can obtain the zero-temperature condition
\be
z-\theta =\frac{2}{3}.
\ee
Near the extremal charged black brane, the Lyapunov exponent $\lambda_n$ of null orbits can be reduced to the form related to $\theta$ or $z$, respectively. We can obtain the formula of $\kappa^2-\lambda_n^2$ in terms of  $\theta$ as
\begin{small}
\be\label{LZTt}
\begin{aligned}
\kappa^2-\lambda_n^2=&\frac{1}{9}r^{2\theta -\frac{16}{3}}\left(48\left(8-10\theta +3\theta^2 \right)+75 r^{\frac{4}{3}}\left(5-8\theta +3\theta^2\right)+24r^{\frac{10}{3}}\left(8-5\theta +3\theta^2 \right)\right.
\\
-&\left.30r^4\left(5-4\theta +3\theta^2 \right) + r^{\frac{20}{3}}\left(9\theta^2-1\right)-40r^{\frac{2}{3}}\left(20-27\theta +9\theta^2 \right) \right),
\end{aligned}
\ee
\end{small}
and the formula in terms of $z$ as
\begin{small}
\be\label{LZTz}
\begin{aligned}
\kappa^2-\lambda_n^2=&\frac{1}{9}r^{2z -\frac{20}{3}}\left(48\left(16-14 z +3z^2 \right)-120 r^{\frac{2}{3}}\left(14-13z+3z^2\right)-30r^{4}\left(9-8z +3z^2 \right)\right.
\\
+&\left.3 r^{\frac{20}{3}}\left(1-4z +3z^2 \right) + 25r^{\frac{4}{3}}\left(35-26z+9^2\right)+8r^{\frac{10}{3}}\left(38-27z +9z^2 \right) \right).
\end{aligned}
\ee
\end{small}

\begin{figure}[htb]
\centering
\subfigure[$\kappa^2-\lambda^2_n$ as the function of $r/r_h$ and $\theta$]{\label{ZTt}
\begin{minipage}[t]{0.48\linewidth}
\centering
\includegraphics[width=1 \textwidth]{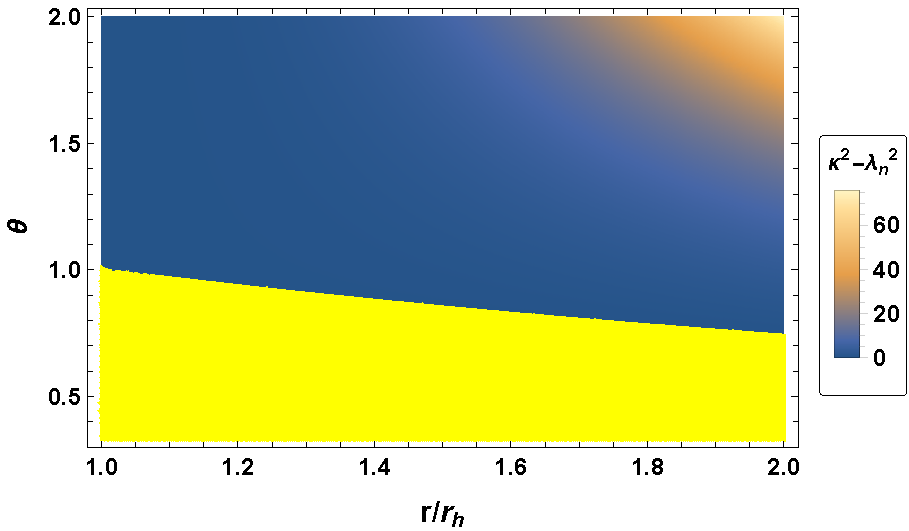}
\end{minipage}
}
\subfigure[$\kappa^2-\lambda^2_n$ as the function of $r/r_h$ and $z$]{\label{ZTz}
\begin{minipage}[t]{0.48\linewidth}
\centering
\includegraphics[width=1 \textwidth]{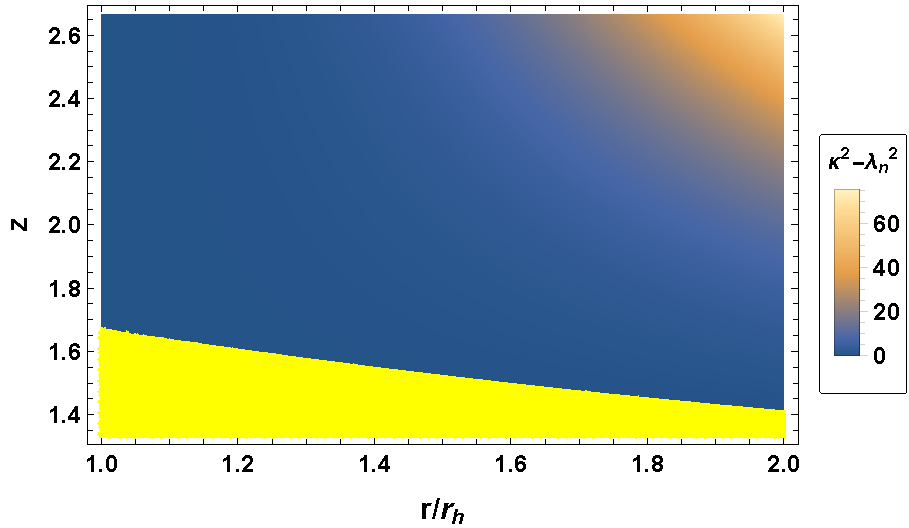}
\end{minipage}
}
\caption{The contour plot of $\kappa^2-\lambda_n^2$ in the extremal charged black brane. In the yellow region, $\kappa^2-\lambda_n^2 <0$, which means the bound $\lambda \leq \kappa$ is violated. }
\label{ZT}
\end{figure}

In \mpref{ZT}, we show the numerical results of $\kappa^2-\lambda_n^2$ in the extremal charged black brane. The region where the chaos bound is violated is colored in yellow. As shown in the two plots, the range of $r/r_h$ for the violation of the bound decreases with the increase of $\theta$ (or $z$) until the chaos bound is no longer violated near the horizon. There are obviously the critical parameter values $\theta_c$ and $z_c$, and when $\theta >\theta_c$ (or $z >z_c$), there is no $\kappa^2 -\lambda_n^2<0$.

To find the critical values of parameters ($\theta$ and $z$), we expand \meqref{LZTt} and \meqref{LZTz} near the horizon ($r_h=1$). There are
\be
\begin{aligned}
\kappa^2-\lambda_n^2 &\sim (\theta -1) (r-1)^3+\mathcal{O}\left((r-1)^4\right),
\end{aligned}
\ee
and
\be
\begin{aligned}
\kappa^2-\lambda_n^2 &\sim (3z-5)(r-1)^3+\mathcal{O}\left((r-1)^4\right).
\end{aligned}
\ee
From the above equations, we can obtain the critical values for the violation of chaos bound in the extremal charged black brane, which we denote here as $\theta_{c1}$ and $z_{c1}$, respectively: $\theta_{c1}=1$ and $z_{c1}=5/3$.

\subsection{$T_{\mu\nu}\xi^\mu\xi^\nu=0$ case}

\begin{figure}[htb]
\centering
\subfigure[$\kappa^2-\lambda^2_n$ as the function of $r/r_h$ and $\theta$]{\label{NECt}
\begin{minipage}[t]{0.48\linewidth}
\centering
\includegraphics[width=1 \textwidth]{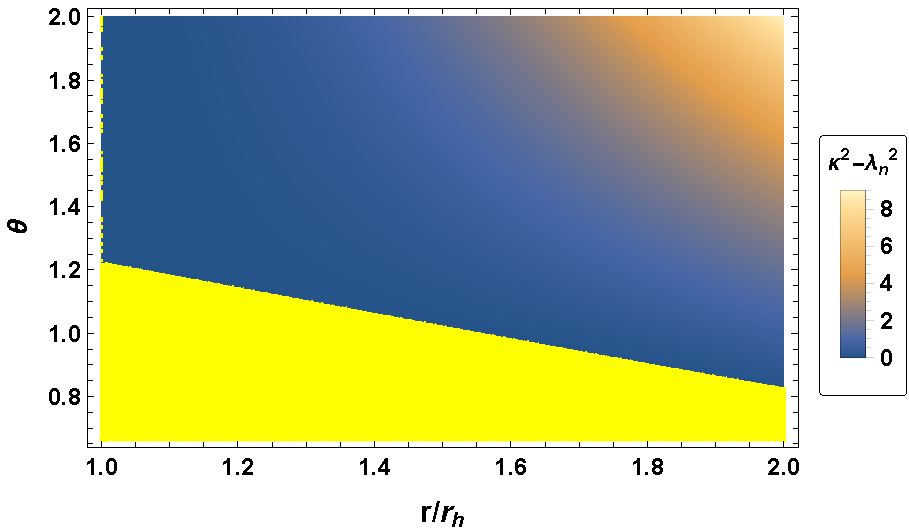}
\end{minipage}
}
\subfigure[$\kappa^2-\lambda^2_n$ as the function of $r/r_h$ and $z$]{\label{NECz}
\begin{minipage}[t]{0.48\linewidth}
\centering
\includegraphics[width=1 \textwidth]{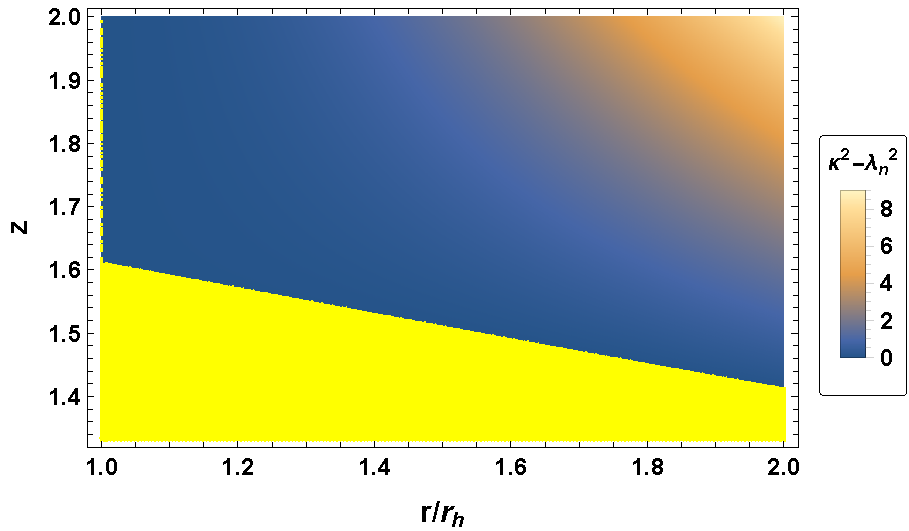}
\end{minipage}
}
\caption{The contour plot of  $\kappa^2-\lambda_n^2$ in the charged black brane with $T_{\mu\nu}\xi^\mu\xi^\nu=0$. In the yellow region, $\kappa^2-\lambda_n^2 <0$, corresponds to the chaos bound $\lambda \leq \kappa$  violated area.}
\label{NEC}
\end{figure}

Next, we discuss the chaos bound at $T_{\mu\nu}\xi^\mu\xi^\nu=0$. In the previous discussion, our results show that $\kappa^2-\lambda^2<0$ is more likely to happen at the parameter range satisfying of $T_{\mu\nu}\xi^\mu\xi^\nu=0$. Here we focus on the charged black brane with the constraint $T_{\mu\nu}\xi^\mu\xi^\nu=0$. With the null energy condition \meqref{necon}, the black brake parameters are constrained by
\be
2z-\theta =2.
\ee
In the critical cases of NEC $T_{\mu\nu}\xi^\mu\xi^\nu=0$, we can simplify the expression of $\kappa^2-\lambda_n^2$ in terms of $\theta$ as
\be\label{LNECt}
\begin{aligned}
\kappa^2-\lambda_n^2=&\frac{1}{16}(2-3\theta)^2+\frac{75}{16}r^{2\theta -4}(\theta -2)^2-15r^{\frac{5\theta}{2}-5}+8r^{3\theta -6}(\theta -2)^2
\\
+&\frac{1}{4}r^{\theta-2}\left(2\theta r^4-5r^{1+\frac{\theta}{2}}(12+\theta (\theta -4))+4r^{\theta} (24+\theta (3\theta -14)) \right),
\end{aligned}
\ee
and the formula of $\kappa^2-\lambda_n^2$ in terms of $\theta$ as
\be\label{LNECz}
\begin{aligned}
\kappa^2-\lambda_n^2=&\frac{1}{4}(4-3z)^2-60r^{5(z-2)}(z-2)^2+32r^{6(z-2)}(z-2)^2+\frac{75}{4}r^{4z-8}(z-2)^2
\\
+ & r^{2z} \left( r^{2z-6} (64+4z(3z-13)) -5r^{z-4} (6+z(z-4))+z-1 \right) .
\end{aligned}
\ee
We show the results of \meqref{LNECt} and \meqref{LNECz} in \mpref{NEC}, and there is $\kappa^2-\lambda_n^2<0$ in the yellow region. For the charged black brane with the critical values of NEC, we can see in \mpref{NEC} that the region of $\kappa^2-\lambda_n^2<0$ in $r/r_h$ shrinks as $\theta$ (or $z$) increases until there is no $\kappa^2-\lambda_n^2$ near the horizon.

Near the horizon $r_h=1$, we can expand \meqref{LNECt} and \meqref{LNECz} as
\be
\begin{aligned}
\kappa^2-\lambda_n^2=&\left(\frac{33}{2}-50\theta +50\theta^2-\frac{81}{4}\theta^3+\frac{99}{32}\theta^4 \right)(r-1)^2+\mathcal{O}(r-1)^3,
\end{aligned}
\ee
and
\be
\begin{aligned}
\kappa^2-\lambda_n^2=&\frac{1}{2}\left(3z-4\right) \left(z\left(394+z\left(33z-196 \right) \right) -264\right)(r-1)^2+\mathcal{O}(r-1)^3.
\end{aligned}
\ee
The critial values $\theta_{c2}$ and $z_{c2}$ for the violation of chaos bound in the cases of $T_{\mu\nu}\xi^\mu\xi^\nu=0$ are $\theta_{c2}=1.219$ and $z_{c2}=1.610$. These values are related to the null energy condition (NEC).

The larger values of $\theta_{c1}$ and $\theta_{c2}$ ($z_{c1}$ and $z_{c2}$) are the critical parameter values: $\theta_c=Max(\theta_{c1},\theta_{c2})$ and $z_c=Max(z_{c1},z_{c2})$. For our setting of parameters ($d=2$, $\phi_0=0$, $Q=2$ and $r_h=1$), the critical parameters are $\theta_c=1.219$ and $z_c=5/3$. In the parameter space $(\theta , z)$ where $\theta>\theta_c$ or $z>z_c$, $\kappa^2-\lambda_n^2$ is always positive, which means the chaos bound is satisfied\footnote{In appendix \ref{VNEC}, we consider the spacetime with the violation of NEC. In the parameter space $(\theta , z)$ larger than the critical parameters $\theta_c$ and $z_c$, the violation of chaos bound still exists. This implies that a deeper physical meaning exists between the chaos bound and the NEC.}.

\section{Conclusion}\label{conclusion}

In summary, we investigate the Lyapunov exponent of a test particle near the charged black branes with the hyperscaling violating factor. The Lyapunov exponent of particle motion near the horizon has an upper bound, which equals the surface gravity $\kappa$ and is called the chaos bound \cite{Hashimoto:2016dfz, Zhao:2018wkl}. The relationship between the equilibrium in the radial direction of test particles and the chaos bound is discussed here. Considering different parameters of the black brane, we find the violation of the chaos bound. Whether the chaos bound can be violated is related to the black brane parameters ($\theta$ and $z$), the black brane temperature $T_H$, and the null energy condition (NEC).

We study the effects of different parameters $\theta$ and $z$ on the particle motion near the black brane, while in our previous work, we pointed out the relationship between the black hole temperature and the chaos bound in the particle motion. The Lyapunov exponent of test particles' equilibrium in the radial direction can be obtained from the Jacobian matrix. The results show that even at the same temperature $T_H$, the test particle still has different behavior of the Lyapunov exponent. Obviously, this is the effect of black brane parameters $\theta$ and $z$. For the equilibrium of test particles, its Lyapunov exponent increases with the lateral momentum. Considering time-like orbits, in the limit where the lateral momentum converges to infinity, the Lyapunov exponent converges to the value of null orbits. Therefore we discuss the Lyapunov exponent $\lambda_n$ of the null orbits and the surface gravity $\kappa$ to study the bound $\lambda \leq \kappa$, since it is most likely to be beyond the bound. We analyze the numerical results of $\kappa^2-\lambda_n^2$ for the same parameters $\theta$ and $z$. The results show that when the hyperscaling violating exponent $\theta$ and the dynamical exponent $z$ are small, there is always $\kappa^2 -\lambda_n^2 <0$. As $\theta$ and $z$ increase, the parameter space $\theta-z$ which has $\kappa^2 - \lambda_n^2<0$ becomes smaller, and $\kappa-\lambda_n^2<0$ only exists in the region near the zero-temperature cases and the cases of $T_{\mu\nu}\xi^\mu\xi^\nu=0$. As the parameters continue to increase, there is no violation of $\lambda \leq \kappa$. These results illustrate that the zero-temperature as well as the violation of NEC gives more possibilities of violating the chaos bound. By considering the limiting examples of zero-temperature and the cases of $T_{\mu\nu}\xi^\mu\xi^\nu=0$, we obtain the critical parameters $\theta_c$ and $z_c$.  The bound $\lambda \leq \kappa$ is always satisfied by the equilibrium in the radial direction of test particles when $\theta>\theta_c$ or $z>z_c$. The critical parameters imply a potential connection between the violation of bound $\lambda \leq \kappa$ and the properties of spacetime. This connection is demonstrated in the Hawking temperature and null energy condition.  Note that the results can recover that of RN-AdS black hole when $z=1$ and $\theta=0$. However, due to the complicated influence of parameters $\theta$ and $z$ on the background spacetime, we do not see  a significant difference in the geometry as the parameters are above or below these critical parameters. The relationship between the Lyapunov exponent's upper bound in particle motion and space-time geometry does require further investigation and exploration.

We obtain the violation of the bound $\lambda \leq \kappa$ in the charged black brane with the hyperscaling violating factor. This result may not be contrary to the conjecture $\lambda \leq 2\pi T/\hbar$ proposed in \cite{Maldacena:2015waa}, since the conjecture was obtained from the out-of-time-order correlator (OTOC) in quantum systems. In holographic theory, it is recognized that OTOC in quantum field theory is dual to the shock wave at the horizon. It is clear that the single particle motion we are discussing is different from the shock wave. At the horizon,  $\lambda =\kappa$ is always satisfied. However, it is still interesting to study the Lyapunov exponent in particle motion and the violation of the bound. In the study on the Lyapunov exponent of particle motion, some interesting conjectures have been proposed. Such as the relationship between the Lyapunov exponent of particle motion, the energy bound \cite{Hashimoto:2021afd} and the causality bound \cite{Hashimoto:2022kfv}. Guo et al. probed the connection between the black hole phase transitions and the Lyapunov exponents of particle motion \cite{Guo:2022kio}. Recently, the relationship between the test particles' homoclinic orbits and chaos bound in the black hole with anisotropic matter fields has also been discussed \cite{Jeong:2023hom}. More about the connection between the chaos bound and the nature of black holes remains to be uncovered.

Our work supports the contact between the violation of the chaos bound and the properties of the black brane (the Hawking temperature and the null energy condition). These results illustrate the potential physical significance of the violation of the chaos bound in particle motion. It will be of great interest to study more about the dynamic stability of black holes through the chaos bound in the particle motion, because there is an intrinsic correlation between them, and figuring out this correlation will help us further understand the black hole.

\section*{Acknowledgement} The work was partially supported by NSFC, China (grant No. 12275166 and No.11875184).

\appendix
\section{The black branes with the violation of NEC}\label{VNEC}
\begin{figure}[htb]
    \centering
    \includegraphics[width=0.80\textwidth]{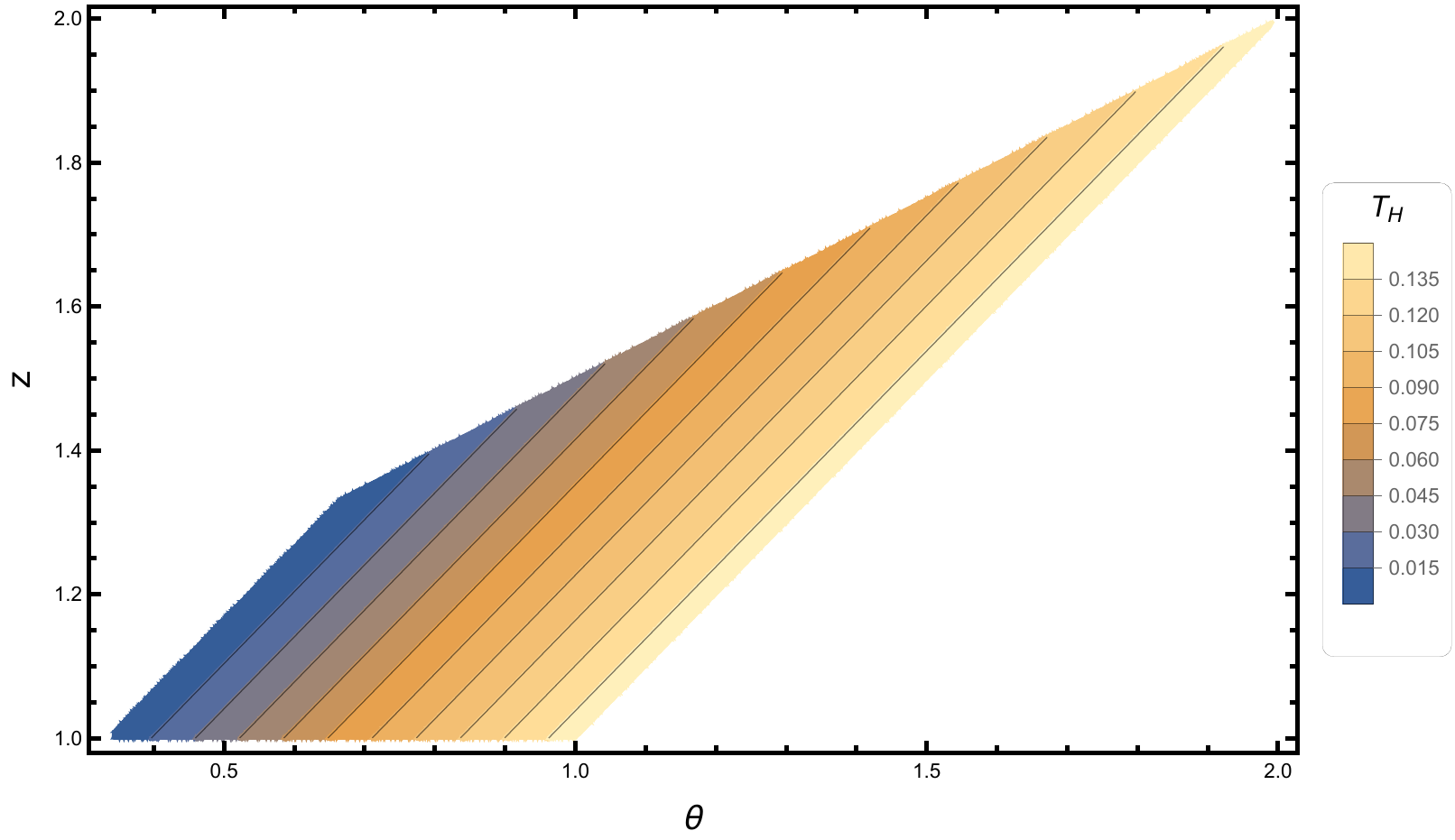}
    \caption{The temperature $T_H$ of the black branes with the violation of NEC as a function of the parameters $\theta$ and $z$.}
    \label{PT2}
\end{figure}

With the parameters
$$
d=2,\qquad \phi_0=0,\qquad Q=2,\qquad r_h=1,
$$
we consider the black branes where NEC is violated. The temperature $T_H$ at the horizon can be obtained by the parameters $\theta$ and $z$, and we show it in \mpref{PT2}. We can see that $T_H$ at the horizon increases as $\theta$ increases and decreases as $z$ increases.

\begin{figure}[!htbp]
\centering
\subfigure[$r/r_h=1.0001$]{\label{R2ra}
\begin{minipage}[t]{0.3\linewidth}
\centering
\includegraphics[width=1\textwidth]{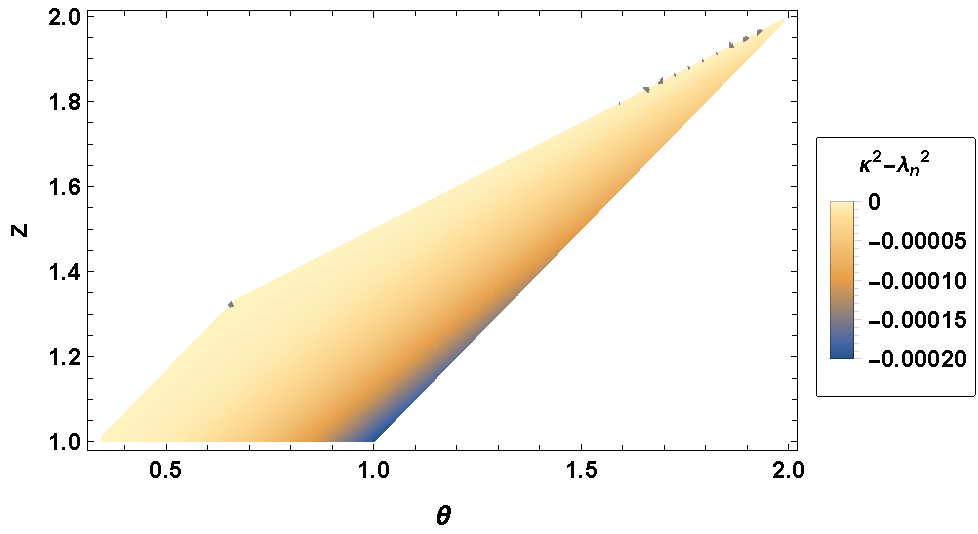}
\end{minipage}
}
\centering
\subfigure[$r/r_h=1.005$]{\label{R2rb}
\begin{minipage}[t]{0.3\linewidth}
\centering
\includegraphics[width=1\textwidth]{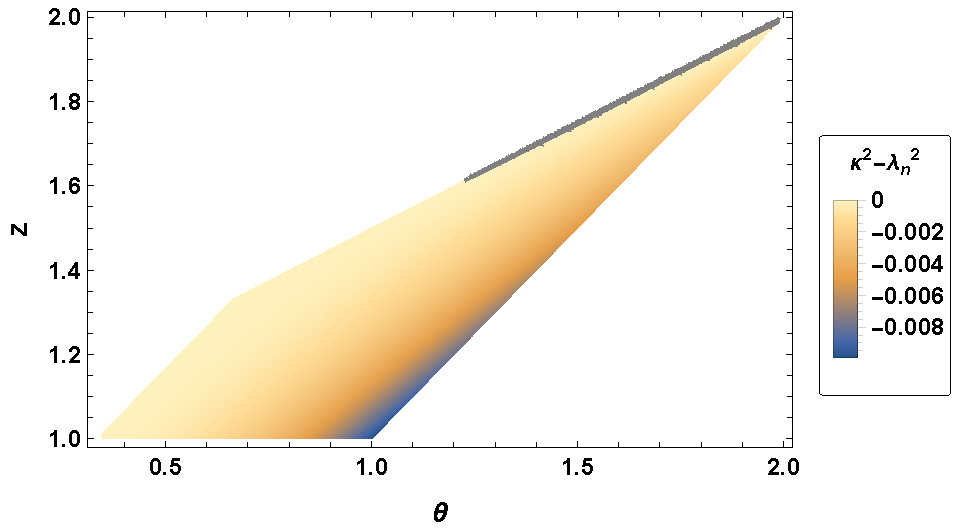}
\end{minipage}
}
\subfigure[$r/r_h=1.1$]{\label{R2rc}
\begin{minipage}[t]{0.3\linewidth}
\centering
\includegraphics[width=1\textwidth]{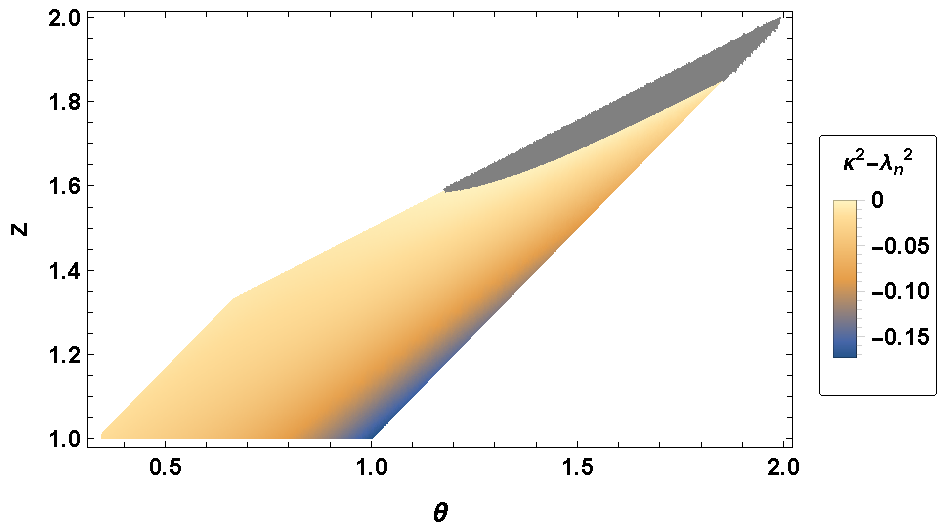}
\end{minipage}
}
\caption{The contour plot of $\kappa^2-\lambda_n^2$ as a function of $\theta$ and $z$ for fixed $r/r_h$: (a) $r/r_h=1.0001$, (b) $r/r_h=1.005$, (c) $r/r_h=1.1$. The colored region in these three plots denotes that $\kappa^2-\lambda_n^2<0$, which means the bound $\lambda \leq \kappa$ is violated. The region which means $\kappa^2-\lambda_n^2>0$ is marked in gray.}
\label{R2r}
\end{figure}

To explore whether the chaos bound can be violated, we compute the value of $\kappa^2-\lambda_n^2$, and if $\kappa^2-\lambda_n^2<0$, it means that the chaos bound is violated. We plot the numerical results of $\kappa^2-\lambda_n^2$ in \mpref{R2r}, which has three plots for different equilibrium positions $r/r_h=1.0001,\ 1.005,\ 1.1$. In the colored region, $\kappa^2-\lambda_n^2<0$, the chaos bound is violated. And the region where $\kappa^2-\lambda_n^2>0$ is marked in gray.

As shown in \mpref{R2r}, in the background where NEC is violated, there is $\kappa^2-\lambda_n^2<0$ in most parameter space $(\theta , z)$. For the equilibrium position $r/r_h=1.0001$ in \mpref{R2ra}, $\kappa^2 - \lambda_n^2<0$ always exists. This result means that in spacetime with the violation of NEC, the chaos bound can always be violated. The results at the equilibrium position $r/r_h=1.005,\ 1.1$ are shown in \mpref{R2rb} and \mpref{R2rc}, respectively. $\kappa^2-\lambda_n^2$ can be positive only when $\theta$ and $z$ are large and close to the cases of $T_{\mu\nu}\xi^\mu\xi^\nu=0$. The closer the equilibrium position is to the horizon, the smaller the range of $\kappa^2-\lambda_n^2>0$. It is worth noting that in black brane with the violation of NEC, the chaos bound is still violated when $\theta >\theta_c$(or $z>z_c$). This implies an intrinsic relationship between the Lyapunov exponent in particle motion and the null energy condition, and it also reminds us of the possible relevance of the chaos bound in particle motion to some interesting questions, such as the instability of spacetime and the causality.


\end{document}